\documentclass[10pt,journal]{IEEEtran}

\ifCLASSOPTIONcompsoc
  \usepackage[nocompress]{cite}
\else
  \usepackage{cite}
\fi

\usepackage{graphicx,subfigure}
\usepackage{amsmath,amsthm,amssymb}
\usepackage{ulem}

\usepackage{xcolor}

\usepackage{algorithm,algorithmic}

\def\F{\mathcal{F}}

\def\P{\Pi \circ}
\newcommand{\vv}[1]{\boldsymbol{#1}} 
\newcommand{\mean}[2]{{\mu}_{{#1}{#2}}} 
\newcommand{\noise}[1]{{\eta}_{#1} }

\usepackage{xcolor}

\title{Modelling Downlink Packet Aggregation in Paced 802.11ac WLANs}
\author{Francesco Gringoli, Douglas J. Leith
\IEEEcompsocitemizethanks{\IEEEcompsocthanksitem F. Gringoli is with University of Brescia, Italy.\IEEEcompsocthanksitem D. J. Leith is with Trinity College Dublin, Ireland. }
}

\begin{document}

\global\csname @topnum\endcsname 0
\global\csname @botnum\endcsname 0
\maketitle

\begin{abstract}
We derive an analytic model of packet aggregation on the the downlink of an 802.11ac WLAN when packet arrivals are paced.  The model is closed-form and so suitable for both analysis and design of next generation edge architectures that aim to achieve high rate and low delay.  The model is validated against both simulations and experimental measurements and found to be remarkably accurate despite its simplicity.
\end{abstract}

\section{Introduction}

In this paper we develop an accurate yet simple analytic model of packet aggregation on the downlink of an 802.11ac WLAN when packet arrivals are paced.  Three factors combine to make finite-load analytic modelling of 802.11ac delay challenging in general: (i) dynamic aggregation of multiple packets in transmitted frames, (ii) randomness in the time between frame transmissions caused by the stochastic nature of the CSMA/CA MAC and (iii) correlated bursty packet arrivals.   As a result most finite-load analysis of packet aggregation to date has resorted to use of simulations.  Use of packet pacing largely mitigates (iii) and we show that this turns out to be sufficient to make accurate analytic modelling tractable.  

On the face of it, packet pacing is a strong assumption.   Use of pacing is becoming more widespread in the internet, e.g. in Google's BBR TCP~\cite{BBR}, but in general in current networks packet arrivals at the downlink of a WLAN can still be expected to be bursty.  Looking ahead, however, proposed next generation architectures for over-the-top services often use a proxy at the network edge (e.g. within a cloudlet) in order to implement new service-driven transport layer behaviours over the path between proxy and user devices, which in particular includes the last wireless hop.  See, for example,~\cite{quickandplenty,ctcp14}.   With such architectures it is straightforward to introduce packet pacing at the network edge and it is this observation that motivates the current work.

Note that one of the most challenging requirements in next generation networks is the provision of connections with low end-to-end latency.  In most use cases the target is for $<$100ms latency, while for some applications it is $<$10ms \cite[Table 1]{ngmn}.  In part, this reflects the fact that low latency is already coming to the fore in network services, but the requirement for low latency also reflects the needs of next generation applications such as augmented reality and the tactile internet.    Use of packet aggregation is ubiquitous in modern WLANs and, as observed for example in \cite{quickandplenty}, the level of packet aggregation is intimately linked to queueing delay at the AP.   This creates a strong need for analytic models of aggregation that can inform next generation low-delay designs.

\section{Related Work}

Interest in using aggregation in WLANs pre-dates the development of the 802.11n standard in 2009 but has primarily focused on analysis and design for wireless efficiency, managing loss etc.  For a recent survey see for example \cite{aggsurvey}.   The literature on throughput modelling of WLANs is extensive but much of it focuses on so-called saturated operation, where transmitters always have a packet to send, see for example \cite{li09} for early work on saturated throughput modelling of 802.11n with aggregation.  When stations are not saturated (so-called finite-load operation) then for WLANs which use aggregation (802.11n and later) most studies resort to the use of simulations to evaluate performance due to the complex interaction between arrivals, queueing and aggregation with CSMA/CA service.   Notable exceptions include \cite{kuppa06,boris09} which adopt a bulk service queueing model that assumes a fixed, constant level of aggregation and \cite{kim08} which extends the finite load approach of \cite{malone07} for 802.11a/b/g but again assumes a fixed level of aggregation.

\section{Modelling Aggregation Level \& Delay}

\subsection{Setup}
We consider downlink transmissions in a WLAN with $n$ client stations indexed by $i=1,2,\dots,n$.   Index the packets arriving at the AP for transmission to station $i$ by $k=1,2,\dots$ and let $\Delta_{i,k}$ denote the inter-arrival time between packet $k-1$ and packet $k$.  The packet sender uses packet pacing, i.e. the sender aims to transmit packets with fixed spacing, although practical constraints typically mean that this aim is only approximately achieved and the packet spacing has some jitter.   We can therefore assume that the $\Delta_{i,k}$ are i.i.d. with $E[\Delta_{i,k}]=\Delta_i=1/x_i$ where $x_i$ is the send rate to station $i$ in packets/sec.   

%
Packets are transmitted to station $i$ by the AP within 802.11 frames.    Each frame may carry multiple packets.  Packets destined for different stations are buffered in separate queues at the AP and these queues are serviced in round-robin fashion.  Namely, when a transmission opportunity occurs the AP takes packets from the next queue $i$ due to be serviced and uses these to form a frame that is then sent to station $i$.   The number of packets sent is determined by the smaller of the queue occupancy and the maximum allowed level $N_{max}$ (typically 64 packets or 5.5ms frame duration, whichever is smallest).    We assume that the packet arrival rate is high enough that the AP transmits at least one packet to each station in every round.   This is reasonable since our primary interest here is in the high rate operation needed for next generation edge-assisted applications. 

Even when packet pacing is used the resulting system exhibits complex behaviour, where the queueing delay of each station is coupled to the packet arrivals of all of the other stations.  To see this observe that the time between transmission opportunities is the sum of two components: (i) the time required to win a channel access using the CSMA/CA protocol and (ii) the time taken to transmit the previous frame.  Due to the stochastic nature of the CSMA/CA protocol the time to win a channel access is random.  The number of packet arrivals between transmission opportunities is therefore also random.   The time taken to transmit a frame depends on the number of packets aggregated in that frame.   Since service is round robin, transmissions to station $i$ are interleaved with transmissions to the other $n-1$ stations.  The time between transmissions to station $i$ therefore depends on the time taken to transmit frames to the other $n-1$ stations, which in turn depends on the number of packets sent in each frame and so on the number of packet arrivals to the other $n-1$ stations.


\subsection{Mean Time Between Frame Transmissions}
The AP transmits frames to the $n$ client stations in a round-robin fashion.  Index rounds by $f=1,2\dots$ and let $\F_{i,f}\subset\{1,2,\dots\}$ denote the set of packets transmitted to station $i$ in the frame sent on round $f$.   Then $N_{i,f}=|\F_{i,f}|$ is the number of packets sent.  Since a minimum of one packet must be contained within a frame and a maximum of $N_{max}$ (typically 32 or 64 packets) then $1\le N_{i,f} \le N_{max}$.   

Suppose, for simplicity, that all packets are of length $l$ bits (this can be easily relaxed).  The airtime used by the frame transmitted to station $i$ in round $f$ is 
\begin{align}%
T_{air,i,f} := T_{oh,i,f}+\frac{l+l_{oh}}{R_{i,f}}N_{i,f} \label{eq:frame}
\end{align}
where $R_{i,f}$ is the PHY data rate used to transmit the payload of the frame, $T_{oh,i,f}$ is the time used for transmission overheads which do not depend on the aggregation level (namely, CSMA/CA channel access, PHY and MAC headers plus transmission of the ACK by the receiver) and $l_{oh}$ is the MAC framing overhead (in bits) for each packet in the frame.     

The duration $\Omega_{f}$ of round $f$ is given by,
\begin{align}
\Omega_{f}&=\sum_{j=1}^nT_{air,j,f}=C_f+\sum_{j=1}^n  \frac{(l+l_{oh})}{R_{j,f}}N_{j,f}
\end{align}
where $C_f=\sum_{j=1}^n T_{oh,j,f}$.   Index stations by the order in which they are serviced by the AP scheduler, i.e. within a round the $i$'th frame transmitted is to station $i$.  In general, the interval $\Omega_{i,f}$ between transmission of frames to station $i$ is not equal to $\Omega_{f}$, but under reasonable assumptions $\Omega_{i,f}$ has the same distribution as $\Omega_{f}$ i.e. $\Omega_{i,f}\sim\Omega_f$.  

In more detail, we have that
\begin{align}
\Omega_{i,f}=&\sum_{j=i}^n (T_{oh,j,f} + \frac{l+l_{oh}}{R_{j,f}}N_{j,f})\notag\\
&+\sum_{j=1}^{i-1} (T_{oh,j,f+1} + \frac{l+l_{oh}}{R_{j,f+1}}N_{j,f+1})\label{eq:round}
\end{align}
The fixed CSMA/CA overhead $T_{oh,j,f}$ associated with channel access etc is i.i.d across stations $j$ and rounds $f$ by virtue of the 802.11 MAC operation (fluctuations in $T_{oh,j,f}$ are due to the CSMA/CA channel access which is uniformly distributed between $0$ and $CW-1$ MAC slots, where $CW$ is the 802.11 contention window).   We therefore have that $\sum_{j=i}^n T_{oh,j,f}+\sum_{j=1}^{i-1} T_{oh,j,f+1}\sim C_f$.   Assume the channel is stationary so that the MCS rate $R_{i,f}$ is identically distributed across rounds $f$ (but of course may vary amongst stations $i$).     Assume also that $N_{i,f}$ and $N_{j,f+1}$ can be approximated as being identically distributed.  It then follows that $\Omega_{i,f}\sim\Omega_f$.

Assuming $N_{i,f}$ and $R_{i,f}$ are independent 
we now have that the mean time between transmissions to station $i$ is
\begin{align}
E[\Omega_{i,f}]=E[\Omega_{f}]&= c+\sum_{j=1}^n\frac{l+l_{oh}}{\mean{R}{_j}}E[N_{j,f}]\notag\\
&=c+\vv{w}^TE[\vv{N}_f]\label{eq:deltai}
\end{align} 
where $c:=E[C_f]=nE[T_{oh,j,f}]$ is the aggregate time used for CSMA/CA channel access, PHY and MAC headers plus transmission of the MAC ACK, $\mean{R}{_i}:=1/E[\frac{1}{R_{i,f}}]$ with $R_{i,f}$ the PHY rate used to transmit frames to station $i$ in round $f$\footnote{Note that in general $E[\frac{1}{R_{i,f}}] \ne 1/E[R_{i,f}]$.  Indeed, to first-order $E[\frac{1}{R_{i,f}}] \approx \frac{1}{E[R_{i,f}] }+ \frac{Var(R_{i,f})}{E[R_{i,f}]^3}$}, $\vv{w}=(\frac{l+l_{oh}}{\mean{R}{_{1}}},\dots,\frac{l+l_{oh}}{\mean{R}{_{n}}})^T$ is the mean time to transmit one packet to each of the $n$ stations, $E[\vv{N}_f]=(E[N_{1,f}],\dots,E[N_{n,f}])^T$ is the mean number of packets transmitted in the frames to the $n$ stations.

\subsection{Mean Number of Packets Sent in Each Frame}
To model the mean number of packets sent in each frame we proceed as follows.   Let $P_{i,f}$ denote the number of packets for station $i$ arriving at the AP during round $f$.   When the time between packets is constant with $\Delta_{i,k}=\Delta_i$ then $P_{i,f}=\Omega_{i,f}/\Delta_i$ (ignoring quantisation effects) but, as already noted, in general we expect some jitter between packet arrivals even when the sender paces its transmissions.   

These packets are buffered in a queue at the AP until they can be transmitted.   Letting $q_{i,f}$ denote the queue occupancy immediately after frame $f$ is transmitted then $q_{i,f+1} = [q_{i,f} + P_{i,f} -N_{i,f+1} ]^+ $.   The AP aggregates as many as possible of these packets queued for transmission into the next frame $f$, thus $N_{i,f}=\min\{q_{i,f} + P_{i,f}, N_{max}\}$ and
\begin{align}
q_{i,f+1} &=[q_{i,f} + P_{i,f}-\min\{q_{i,f} + P_{i,f}, N_{max}\}]^+ \\
&=[q_{i,f} + P_{i,f} -N_{max} ]^+\label{eq:q}
\end{align}
There are three operating regimes to consider:
\begin{enumerate}
\item Firstly, when $E[P_{i,f}] > N_{max}$ then the queue is unstable.  The queue occupancy grows and so $N_{i,f+1}=N_{max}$ eventually for all frames $f$.   This corresponds to saturated operation and has been studied elsewhere (and is not of interest in the present work where our focus is in the finite-load regime).
\item Secondly, our main interest is in the regime where the queue backlog remains low {i.e.} $P_{i,f}<N_{max}$.  The queue is cleared by each frame transmission so $q_{i,f+1}=0$ and $N_{i,f+1}=P_{i,f}$.   
\item Thirdly, there is the transition regime between regimes one and two where $E[P_{i,f}] < N_{max}$ but $P_{i,f}$ may sometimes be greater than $N_{max}$ and $q_{i,f+1}$ may be non-zero.   
\end{enumerate}
In regime two, $N_{i,f+1}=P_{i,f}$.  Taking expectations $E[N_{i,f+1}]=E[P_{i,f}]=E[\Omega_{i,f}]/E[\Delta_{i,k}]$ (by renewal-reward theory since the $\Delta_{i,k}$ are i.i.d and independent of $\Omega_{i,f}$).   Let $\mean{\vv{N}}{}=(\mean{N}{_1},\dots,\mean{N}{_n})^T$ with $\mean{N}{_i}:=E[N_{i,f+1}]$, and recall that $x_i:=1/E[\Delta_{i,k}]=1/\Delta_i$ is the send rate of station $i$.  Then  substituting from (\ref{eq:deltai}) it follows that $\mean{\vv{N}}{}= ( c +  \vv{w}^T\mean{\vv{N}}{})\vv{x}$.  Rearranging yields  
\begin{align}
\mean{\vv{N}}{}=\frac{c\vv{x}}{1-\vv{w}^T\vv{x}}\label{eq:N}
\end{align}
where $\vv{x}=(x_1,\dots,x_n)^T$ is the vector of station send rates.   

To simplify the analysis we assume that the third regime can be lumped with regime two\footnote{Our measurements in Section \ref{sec:val} support the validity of this simplifying assumption.  In practice it amounts to assuming that the system transitions quickly between operating regimes one and two, i.e. regime three is only transient.  
}.  In regime three $E[P_{i,f}]>N_{max}$ and $E[N_{i,f+1}]=N_{max}$.  Incorporating the $N_{max}$ constraint into (\ref{eq:N}) gives the following expression for the mean number of packets sent in each frame in regime two,
\begin{align}
\mean{\vv{N}}{}=\P\frac{c\vv{x}}{1-\vv{w}^T\vv{x}} =\P \vv{F}(\vv{x})\label{eq:model}
\end{align}
where $\Pi$ denotes projection onto interval $[1,N_{max}]$ and $\vv{F}(\vv{x}):=\frac{c\vv{x}}{1-\vv{w}^T\vv{x}}$.  


\subsection{Fluctuations Around Mean} 
Equation (\ref{eq:N}) models the relationship between the arrival rate $\vv{x}$ and the mean aggregation level $\mean{\vv{N}}{}$ when operating in regime two.   We can also obtain an approximate model of the fluctuations $\noise{{N}_{i,f}}=N_{i,f}-\mean{N}{_{i}}$ about the mean as follows.   Neglecting the jitter in the packet inter-arrival times then the number of packets $P_{i,f}$ arriving at the AP during round $f$ is approximately $\Omega_{i,f}x_i$.   That is, the fluctuations in $P_{i,f}$ (and so the number $N_{i,f+1}$) of packets transmitted in a frame are induced by fluctuations in the duration $\Omega_{i,f}$ of the scheduling round for station $i$.  Neglecting the impact of the position of each station within a round then $\Omega_{i,f}\approx \Omega_f$ (this is exact in the case of a single station).   Combining these we obtain the model
\begin{align}
\vv{N}_{f+1}=(C_f+\vv{w}^T\vv{N}_f)\vv{x}
\end{align}
Since $\mean{\vv{N}}{}= ( c +  \vv{w}^T\mean{\vv{N}}{})\vv{x}$ it follows that
\begin{align}
\noise{\vv{N}_{f+1}}=\vv{x}\vv{w}^T\noise{\vv{N}_{f}}+(C_f-c)\vv{x}\label{eq:eta}
\end{align}
where $\noise{\vv{N}_{f}}=[\noise{{N}_{1,f}},\dots,\noise{{N}_{n,f}}]^T$.  Observe that $\noise{\vv{N}_{f}}$ evolves according to first-order dynamics driven by i.i.d stochastic input $(C_f-c)\vv{x}$.   In 802.11ac $C_f-c$ is a random variable uniformly distributed between 0 and 135$\mu $s (CWmin is 16 and a MAC slot is $9\mu$s).  The matrix $\vv{x}\vv{w}^T$ is rank one and has one zero eigenvalue $\vv{w}^T\vv{x}=\sum_{i=1}^nw_ix_i=\sum_{i=1}^n(l+l_{oh})x_i/\mean{R}{_i}$ and an eigenvalue of zero with multiplicity $n-1$.    The time constant of the dynamics is therefore $\tau:-=E[\Omega_f]/\log(\vv{w}^T\vv{x})$.   

\subsection{Upper Bound On Mean Queueing Delay}

With expression (\ref{eq:model}) for the mean number of packets sent in each frame in hand we are now in a position to model the mean queueing delay.  In operating regime two, the queue is cleared after each transmission.  Hence, the first packet sent to station $i$ in round $f$ arrives to an empty queue and must wait $\sum_{k\in\F_{i,f}}\Delta_{i,k}$ seconds before the last packet sent arrives at the AP and so becomes available for transmission.  The delay experienced by the first packet (and so by all other packets sharing the frame sent in round $f$) is at most $\sum_{k\in\F_{i,f}}\Delta_{i,k}$.  This upper bound is attained if the frame is transmitted right before arrival of the first packet sent in the next round $f+1$ since if the frame in round $f$ was transmitted later then this packet would have been added to frame $f$.  That is, mean delay of packets sent to station $i$ is upper bounded by,
\begin{align}
\mean{T}{_i} &= E[\sum_{k\in\F_{i,f}}\Delta_{i,k}]=E[N_{i,f}]\Delta_i=\frac{\mean{N}{_i}}{x_i}\\
&=\max\{\min\{\frac{c}{1-\vv{w}^T\vv{x}},\frac{N_{max}}{x_i}\},\frac{1}{x_i}\}\label{eq:delay}
\end{align}
\section{Model Validation}\label{sec:val}
\subsection{NS3 Simulator Implementation}\label{sec:ns3}
We  implemented a paced sender in the NS3 packet-level simulator.  Based on the received feedbacks it periodically configures the sending rate of {\tt udp-client} applications colocated at a single node connected to an Access Point. Each wireless station receives a single UDP traffic flow at a {\tt udp-server} application that we modified to collect frame aggregation statistics.  We also developed a round-robin scheduler at the AP with separate queue for each destination, and we added new functions to let stations determine the MCS of each received frame together with the number of MPDU packets it contains.  The maximum aggregation level permitted is $N_{max}$=64.  We configured 802.11ac to use a physical layer operating over an $80MHz$ channel, VHT rates for data frames and legacy rates for control frames.  The PHY MCS and the number of spatial streams NSS used can be adjusted.  As validation we reproduced a number of the simulation measurements in our experimental testbed and found them to be in good agreement. {The new NS3 code and the software that we used to perform experimental evaluations are available open-source}\footnote{Code can be obtained by contacting the corresponding author.}. 

\subsection{Experimental Testbed}
Our experimental testbed uses an Asus RT-AC86U Access Point (which uses a Broadcom 4366E chipset and supports 802.11ac MIMO with up to three spatial streams.   It is configured to use the 5GHz frequency band with 80MHz channel bandwidth.   This setup allows high spatial usage (we observe that almost always three spatial streams are used) and high data rates (up to MCS 9).    By default aggregation supports AMSDU's and allows up to 128 packets to be aggregated in a frame (namely 64 AMSDUs each containing two packets).  A Linux server connected to this AP via a gigabit switch uses iperf 2.0.5 to generate UDP downlink traffic to the WLAN clients.     Iperf inserts a sender-side timestamp into the packet payload and since the various machines are tightly synchronised over a LAN this can be used to estimate the one-way packet delay (the time between when a packet is passed into the socket in the sender and when it is received).  Note, however, that in production networks accurate measurement of one-way delay is typically not straightforward as it is difficult to maintain accurate synchronisation between server and client clocks (NTP typically only synchronises clocks to within a few tens of milliseconds).  

\begin{figure}
\centering
\subfigure[One and two stations (same send rate), MCS=9, NSS=2, NS3]{
\includegraphics[width=0.46\columnwidth]{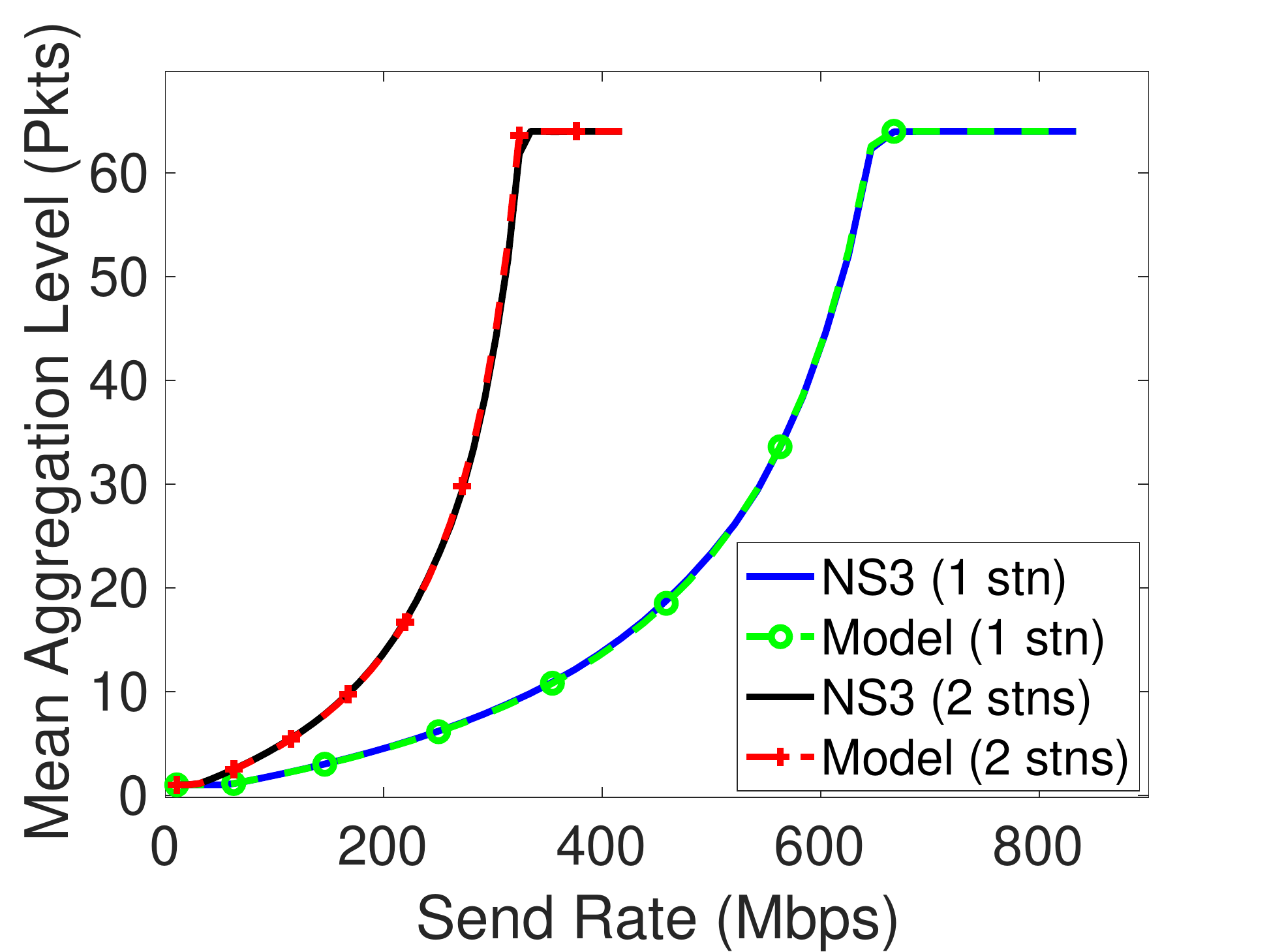}
}
\subfigure[One and two stations (same send rate), testbed data]{
\includegraphics[width=0.46\columnwidth]{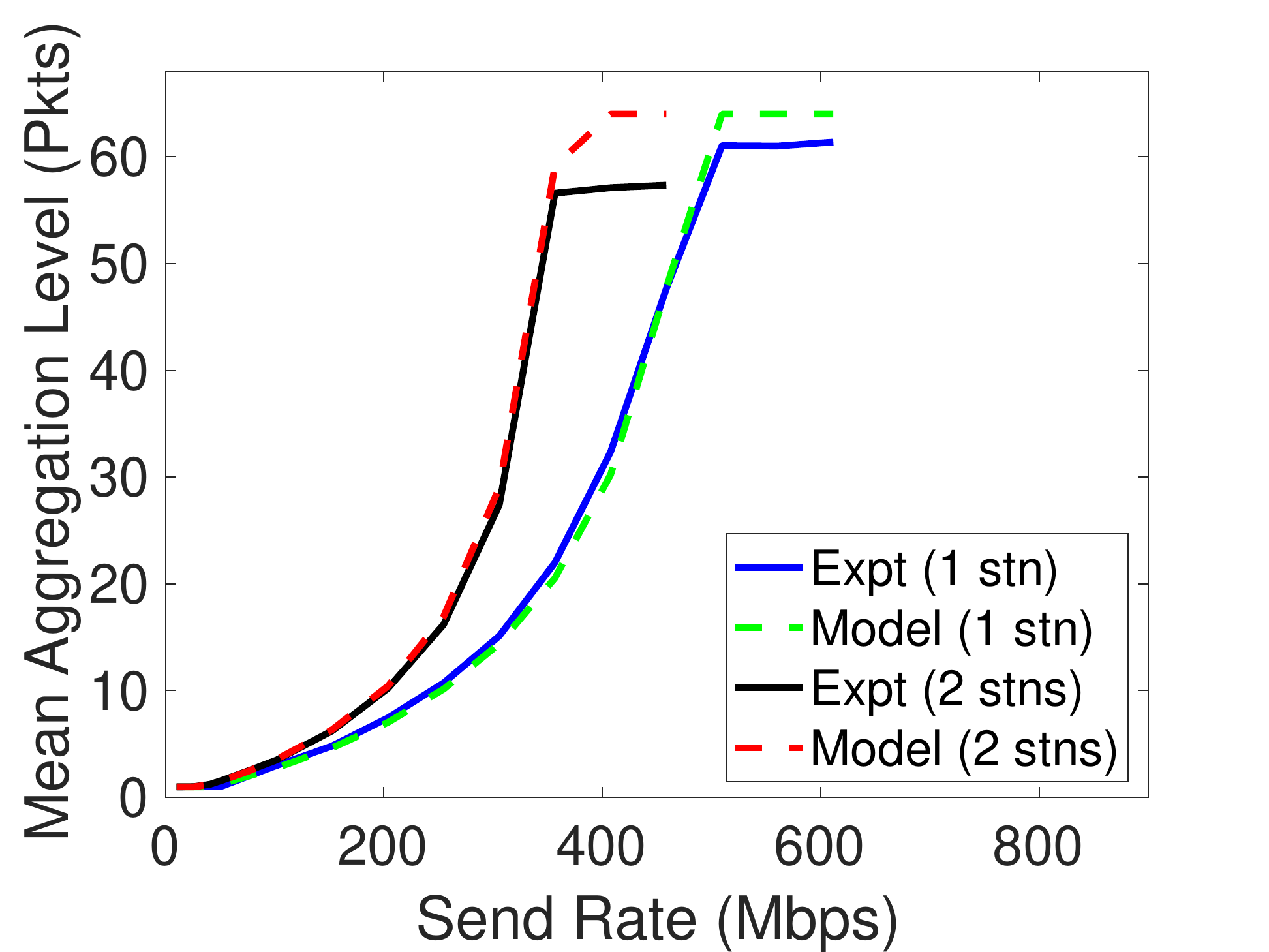}
}
\subfigure[One station, MCS and send rate varied, NSS=3, NS3]{
\includegraphics[width=0.46\columnwidth]{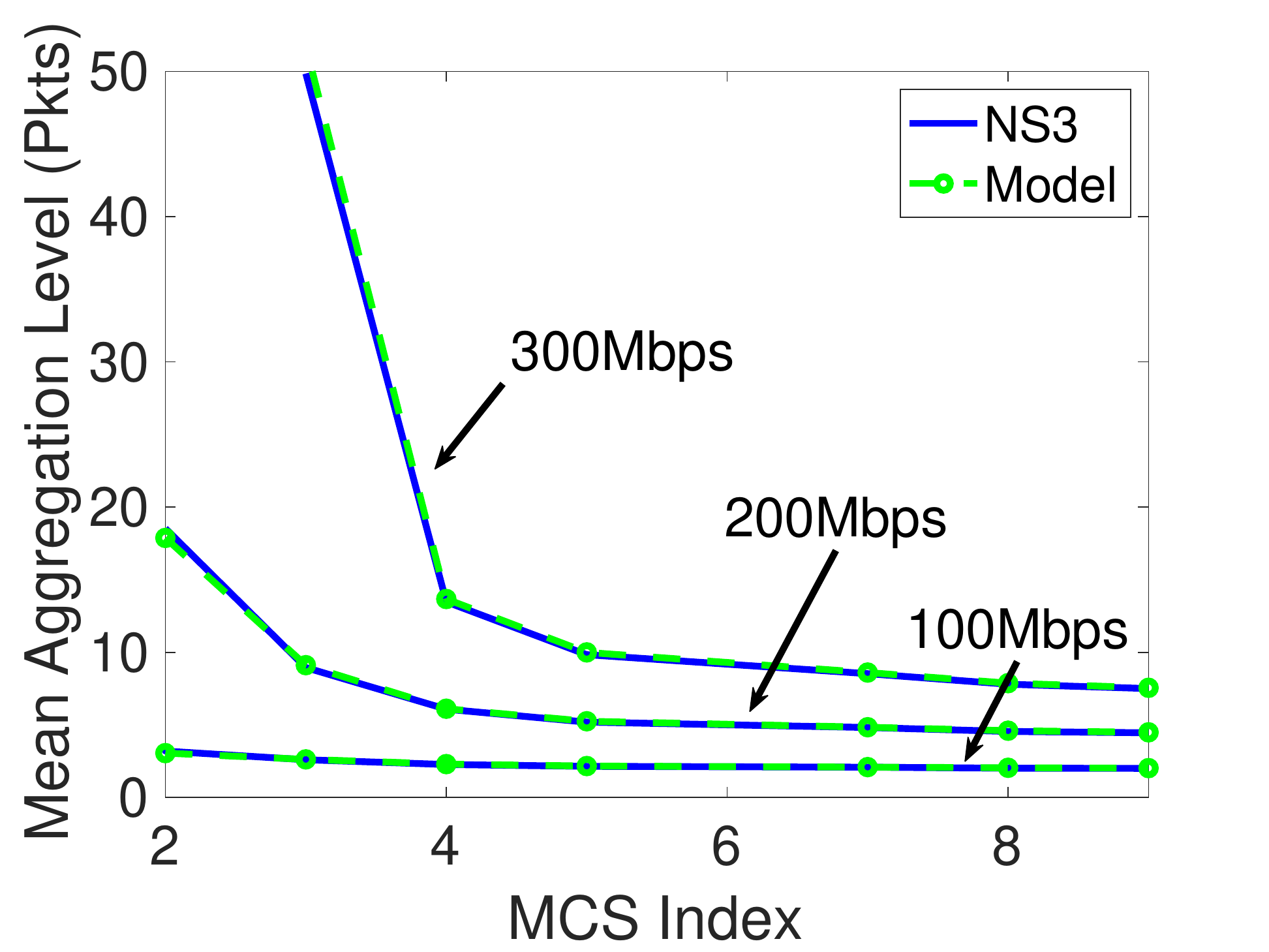}
}
\subfigure[Two stations with different send rates and MCSs.  MCS 9 for station 1, MCS 3 for station 2, NSS=1, NS3]{
\includegraphics[width=0.46\columnwidth]{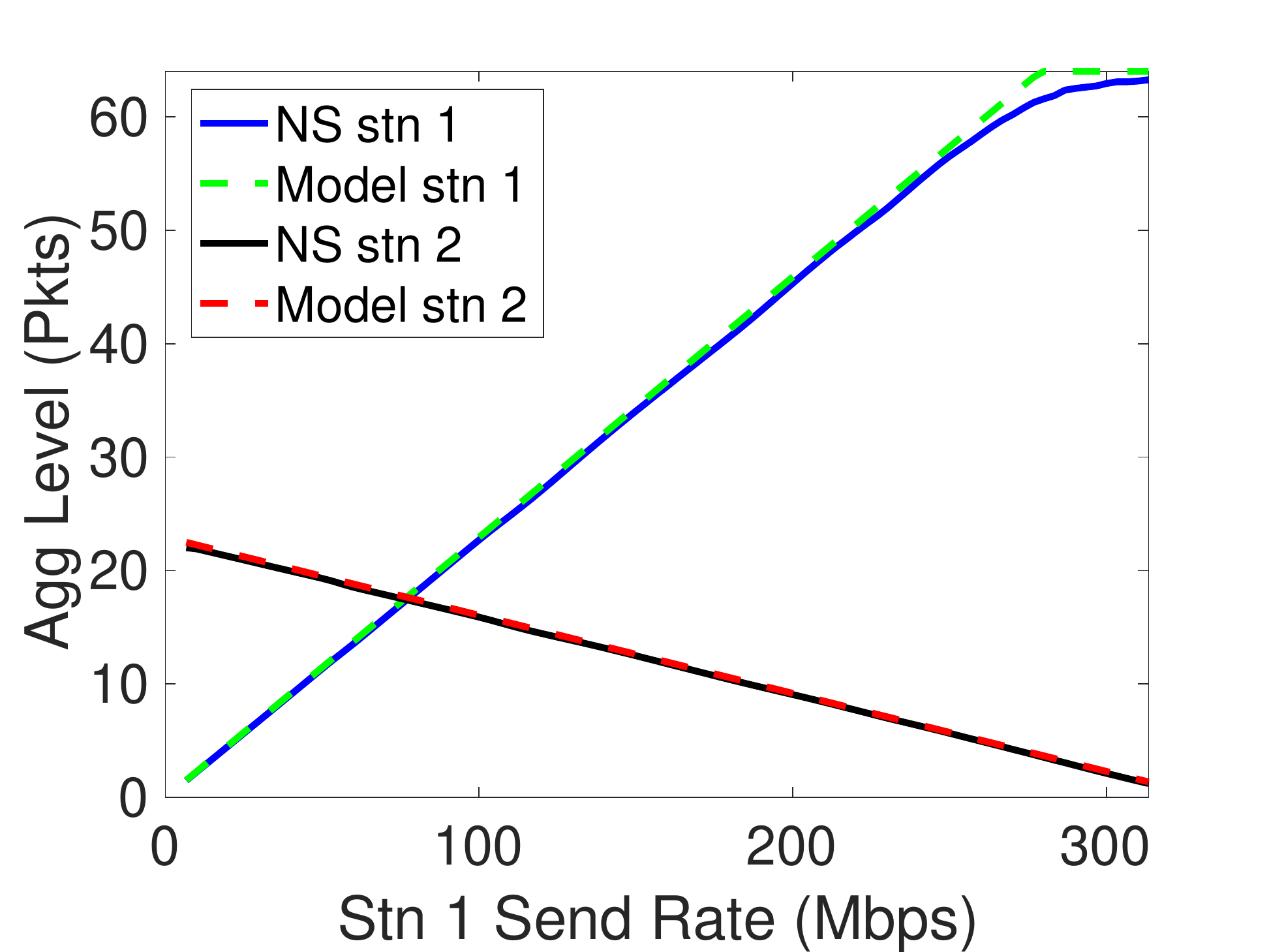}
}
\caption{Comparison of model (\ref{eq:model}) with measurements .    Data is shown for sending UDP packets to one and two client stations, the same send rate being used to all stations and indicated on the x-axes of the plots.  Plots (a),(c),(d) compare the mean aggregation level measured from NS3 simulations with the model, plot (b) compares measurements from an experimental testbed.  In (a)-(c) when there are two stations they have the same send rate, in (d) the stations have different send rates: the send rate to station 1 increases from 5 to 310Mbps while the send rate to station 2 decreases from 100 to 5Mbps.}\label{fig:agg_theory}
\end{figure}


\subsection{Validation of Mean Aggregation Level}
In Figure \ref{fig:agg_theory}(a) we compare the predictions of model (\ref{eq:model}) against measurements from the NS3 detailed packet level simulator as the arrival rate $\vv{x}$ is varied.    Data is shown for the case of a single client station and when there are two client stations both with the same arrival rate.   The values of the model parameters $c$ and $w$ are derived from the 802.11ac MAC/PHY settings (80MHz channel, MCS 9, NSS 2).  It can be see that the model is in remarkably good agreement with the simulation data.    We also collected measurements of aggregation level vs arrival rate in our experimental testbed.  Figure \ref{fig:agg_theory}(b) compares these experimental measurements against the model predictions\footnote{802.11ac settings: NSS=3, 80Mhz channel, the MCS used fluctuates over time due to the action of the 802.11ac rate controller and so an average value is used.  The model $c$ and $w$ parameter values used in Figure \ref{fig:agg_theory}(b) are $c=270\mu s$, $\mean{R}{}=585$Mbps for the one station data and $c=320\mu s$, $\mean{R}{}=850$Mbps for the two station data.}  and again it can be seen that there is excellent agreement between the model and the measurements.

The model (\ref{eq:model}) predicts that the aggregation level scales as the reciprocal of $1-\vv{w}^T\vv{x}=1-\sum_{i=1}^nL/\mean{R}{_i}$.  Figure \ref{fig:agg_theory}(c) compares the model predictions as the MCS rate $\mean{R}{_i}$ is varied (for the 802.11ac setting used $\mean{R}{_i}= 87.8$Mbps at MCS index 0 increasing to $1170$Mbps at MCS index 9).   The model also predicts that for the ratio of the aggregation level of two stations is proportional to the ratio of their send rates and this behaviour is evident in Figure \ref{fig:agg_theory}(d) which plots the aggregation level for two stations when the send rate to the first station is increased from 5 to 310Mbps while that to the second station is decreased from 100 to 5Mbps.

In summary, the model (\ref{eq:model})  is in good agreement with measurements with regard to the dependence of aggregation level on overall send rate, MCS and ratio of station send rates.

\subsection{Validation of Fluctuations Around Mean Aggregation Level}
\begin{figure}
\centering
\subfigure[]{
\includegraphics[width=0.46\columnwidth]{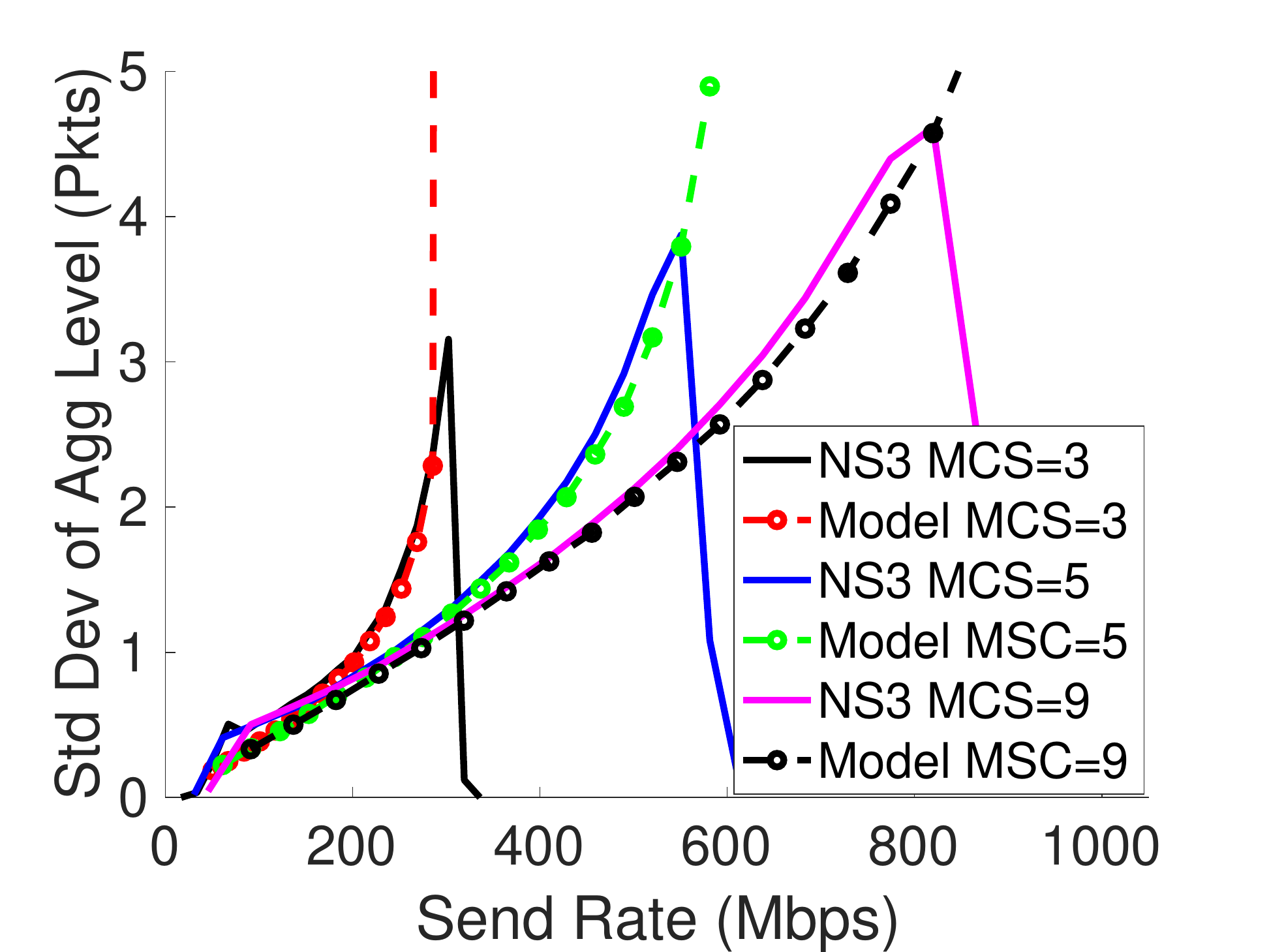}
}
\subfigure[]{
\includegraphics[width=0.46\columnwidth]{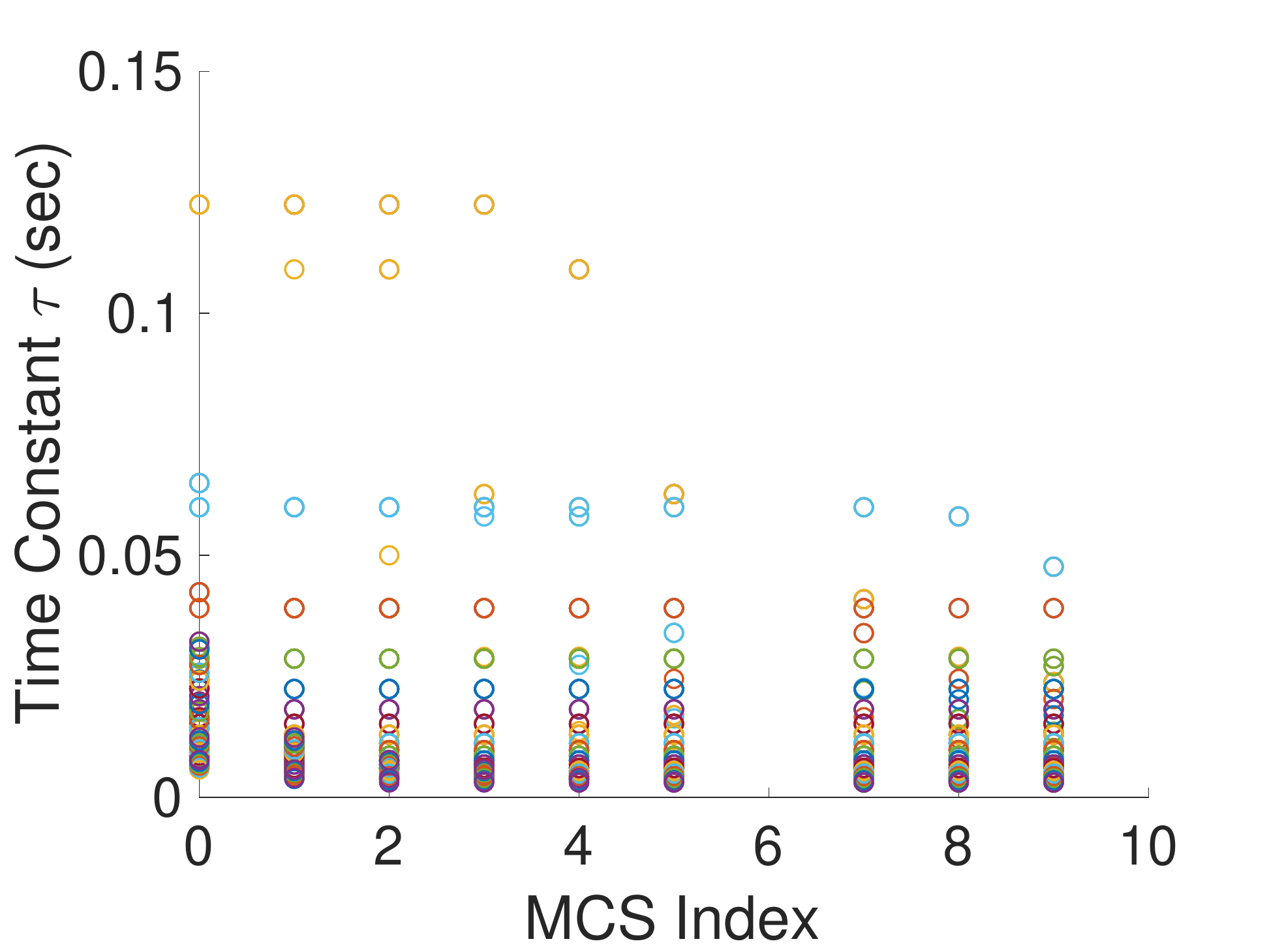}
}
\subfigure[]{
\includegraphics[width=0.46\columnwidth]{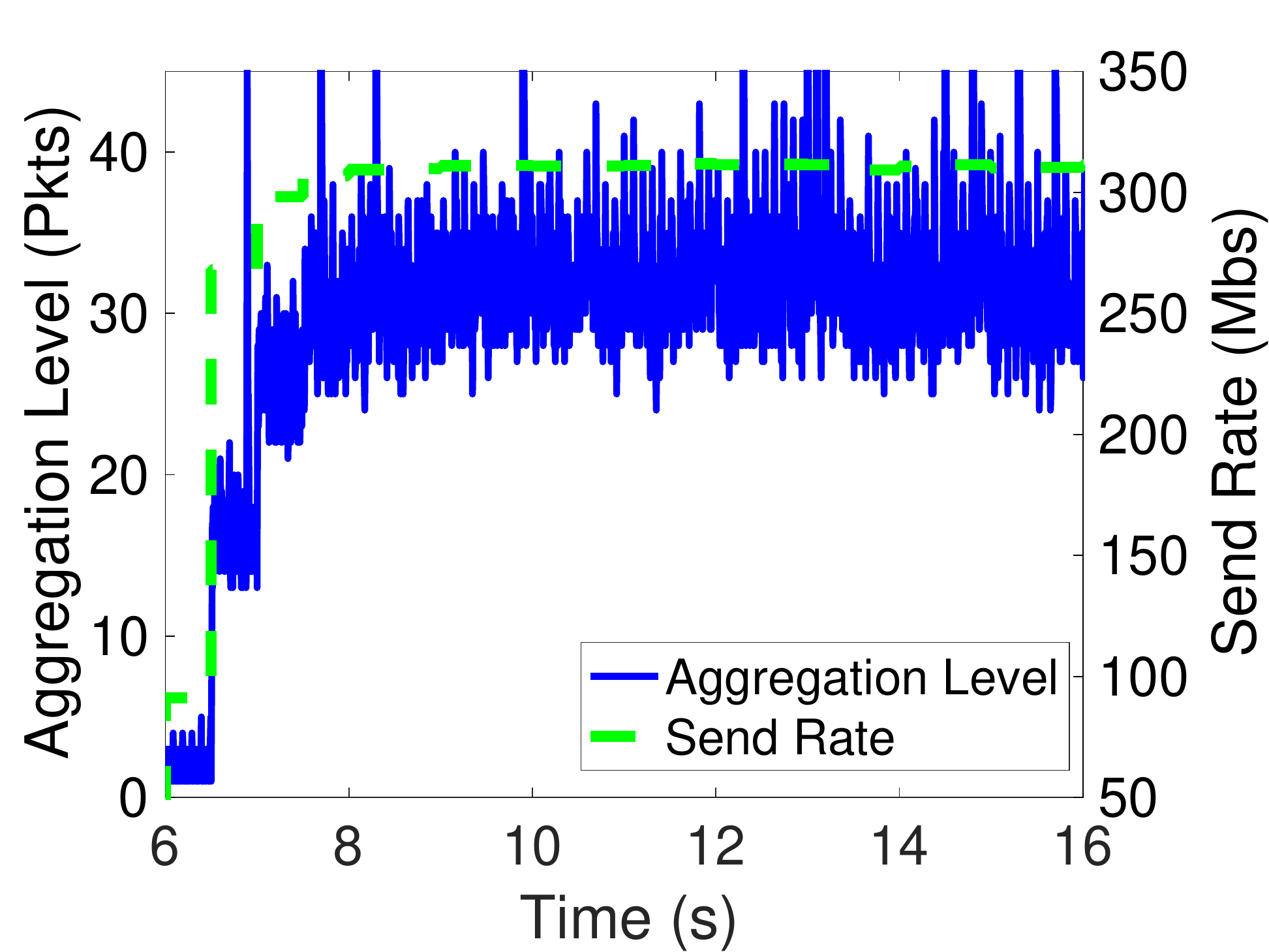}
}
\subfigure[]{
\includegraphics[width=0.46\columnwidth]{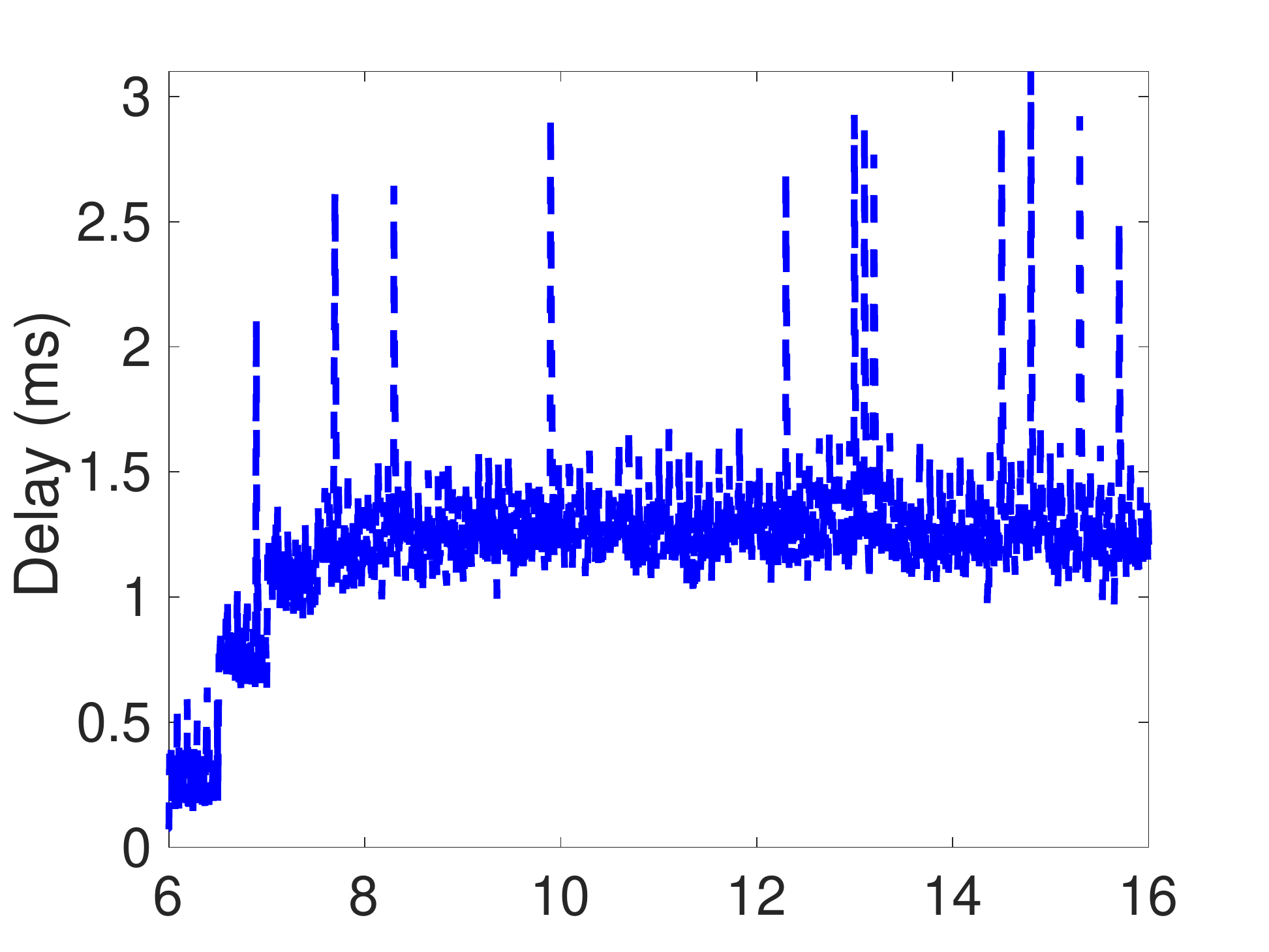}
}

\caption{(a) Comparison of the standard deviation of $\noise{\vv{N}_{f}}$ calculated using the model (\ref{eq:eta}) with measurements from NS3, NSS=3.  (b) time constant of dynamics (\ref{eq:eta}) as the number $n$ of stations is varied between 1 and 20, NSS is varied from 1 to 3 and the MCS index is varied fro 0 to 9 i.e. covering the full 802.11ac NSS/MCS range.  (c) and (d) show example time histories of frame aggregation level and packet delay as send rate is varied with MCS 9, NSS 1.} \label{fig:eta}
\end{figure}

Figure \ref{fig:eta}(a) compares the predictions of the standard deviation of $\noise{\vv{N}_{f}}$ calculated using the model (\ref{eq:eta}) with measurements of the standard deviation of the aggregation level from NS3.  Data is shown as the send rate and MCS rate are varied.  It can be seen that the model predictions are in good agreement with the measurements except when the aggregation level hits its maximum value $N_{max}$, at which point the standard deviation of the measured data falls sharply to zero.   That is, the model (\ref{eq:eta}) is accurate within operating regime two but not in operating regime three, as expected.

Observe that the standard deviation of $\noise{\vv{N}_{f}}$ increases with the send rate, which is intuitive.  The main source of the fluctuations in $\vv{N}_{f}$ is the randomness in the channel access time associated with the operation of the CSMA/CA MAC.   During a round where the channel access randomness leads to the round being of longer than average duration then more  packets arrive than on average, with the number arriving increasing with the send rate.  At the next round these packets form the next frame, which is therefore larger than average.   The magnitude of the fluctuations $\noise{\vv{N}_{f}}$ in the frame size therefore tends to increase with the send rate.

Note that larger frames also tends to make the next round longer than average since they take longer than average to transmit.   This creates feedback whereby a random fluctuation in the duration of a round tends to create changes that persist for several rounds.   It is this feedback that is reflected in the dynamics (\ref{eq:eta}).  

The measurement data in Figure \ref{fig:eta}(a) includes packet inter-arrival jitter of $\pm 6\mu$s.  We also collected measurements for other values of jitter and found the standard deviation of $\noise{\vv{N}_{f}}$ to be largely insensitive to the level of pacing jitter.

%
%

Figure \ref{fig:eta}(b) plots the value of the time constant $\tau$ of the dynamics as the number of stations is varied from 1 to 20, NSS is varied from 1 to 3 and the MCS index from 0 to 9.  For each configuration the aggregation level $N$ is the minimum of $N_{max}$ and the level for which the mean delay $\mean{\vv{T}}{}$ is 5ms.  It can be seen that the time constant is never more than about 0.12s, and tends to fall with increasing MCS rate.   

Figure \ref{fig:eta}(c) shows a typical time history as the send rate is increased in steps.    Observe that that the magnitude of the fluctuations varies with the send rate e.g they are significantly lower in the early part of the time history around 6-6.5s, where the send rate is lower, than from 10s onwards.    Figure \ref{fig:eta}(d) shows the corresponding packet delay time history and, as expected, it can be seen that the delay behaviour essentially mimics the aggregation level.

\section{Conclusions}
We derive an analytic model of packet aggregation on the the downlink of an 802.11ac WLAN when packet arrivals are paced.  The model is closed-form and so suitable for both analysis and design of next generation edge architectures that aim to achieve high rate and low delay.  The model is validated against both simulations and experimental measurements and found to be remarkably accurate despite its simplicity.

\bibliographystyle{IEEEtran}
\bibliography{references.bib}

\begin{IEEEbiography}{Francesco Gringoli} received  the  Laurea  degree  in  telecommunications  engineering  from the  University  of  Padua,  Italy,  in  1998  and  the PhD degree  in  information  engineering  from the  University  of  Brescia,  Italy,  in  2002.  Since 2018 he is Associate Professor  of Telecommunications at the Dept. of Information Engineering at the University of Brescia, Italy. His research interests include security assessment, performance evaluation and medium access control in Wireless LANs. He is a senior member of the IEEE.
\end{IEEEbiography}

\begin{IEEEbiography}{Doug Leith} graduated from the University of Glasgow in 1986 and was awarded his PhD, also from the University of Glasgow, in 1989. In 2001, Prof. Leith moved to the National University of Ireland, Maynooth and then in Dec 2014 to Trinity College Dublin to take up the Chair of Computer Systems in the School of Computer Science and Statistics.  His current research interests include wireless networks, network congestion control, distributed optimization and data privacy.  
\end{IEEEbiography}
\end{document}